\documentclass[sigconf]{acmart}

\AtBeginDocument{%
  \providecommand\BibTeX{{%
    \normalfont B\kern-0.5em{\scshape i\kern-0.25em b}\kern-0.8em\TeX}}}

\usepackage{wrapfig}

\usepackage[most]{tcolorbox}
\usepackage{xcolor}

\newcommand{\camera}[1]{{\color{black}#1}} 
\newcommand{\revise}[1]{{\color{black}#1}}

\copyrightyear{2026}
\acmYear{2026}
\setcopyright{cc}
\setcctype{by-nc-nd}
\acmConference[DIS '26]{Designing Interactive Systems Conference}{June 13--17, 2026}{Singapore, Singapore}
\acmBooktitle{Designing Interactive Systems Conference (DIS '26), June 13--17, 2026, Singapore, Singapore}
\acmDOI{10.1145/3800645.3812880}
\acmISBN{979-8-4007-2563-0/2026/06}

\usepackage{graphicx}

\begin{document}

\title  {Otherness as a Quality in Designing Expressive Robotic Touch}

%

\author{Ran Zhou}
\affiliation{%
  \institution{KTH Royal Institute of Technology}
  \city{Stockholm}
  \country{Sweden}}
\email{ranzhou@kth.se}

\author{Laurens Boer}
\affiliation{%
  \institution{IT University of Copenhagen}
  \city{Copenhagen}
  \country{Denmark}}
\email{laub@itu.dk}

\author{Daniel Leithinger}
\affiliation{%
  \institution{Cornell University}
  \city{Ithaca}
  \state{New York}
  \country{USA}}
\email{daniel.leithinger@cornell.edu}

\author{Madeline Balaam}
\affiliation{%
  \institution{KTH Royal Institute of Technology}
  \city{Stockholm}
  \country{Sweden}}
\email{balaam@kth.se}



\renewcommand{\shortauthors}{Zhou et al.}

\begin{abstract}

Haptic technologies have advanced rapidly, yet exploration of robotic touch remains dominated by replicating realistic environmental cues or hand gestures, which narrows the design space and risks social resistance. This paper argues for alternatives: grounded in the notion of “otherness” from human–robot interaction \camera{(HRI)}, we propose treating robotic touch’s inherent otherness as a design quality. Instead of being a limitation in pursuing realism, otherness can be embraced to elicit ambiguity and provoke alternative interpretations, fostering expressive and evocative robotic touch design. To develop this perspective, \revise{we analyze inspirational art and design precedents and four design research cases through a reflective Research through Design \camera{(RtD)} approach. Through this analysis, we articulate a set of design languages structured around why otherness matters for touch meaning-making, how it can be shaped through design strategies, and where it can be embedded within robotic touch systems. We conclude by reflecting on the tensions and risks involved in designing robotic touch with otherness in mind.}



\end{abstract}



\begin{teaserfigure}
  \includegraphics[width=\textwidth]{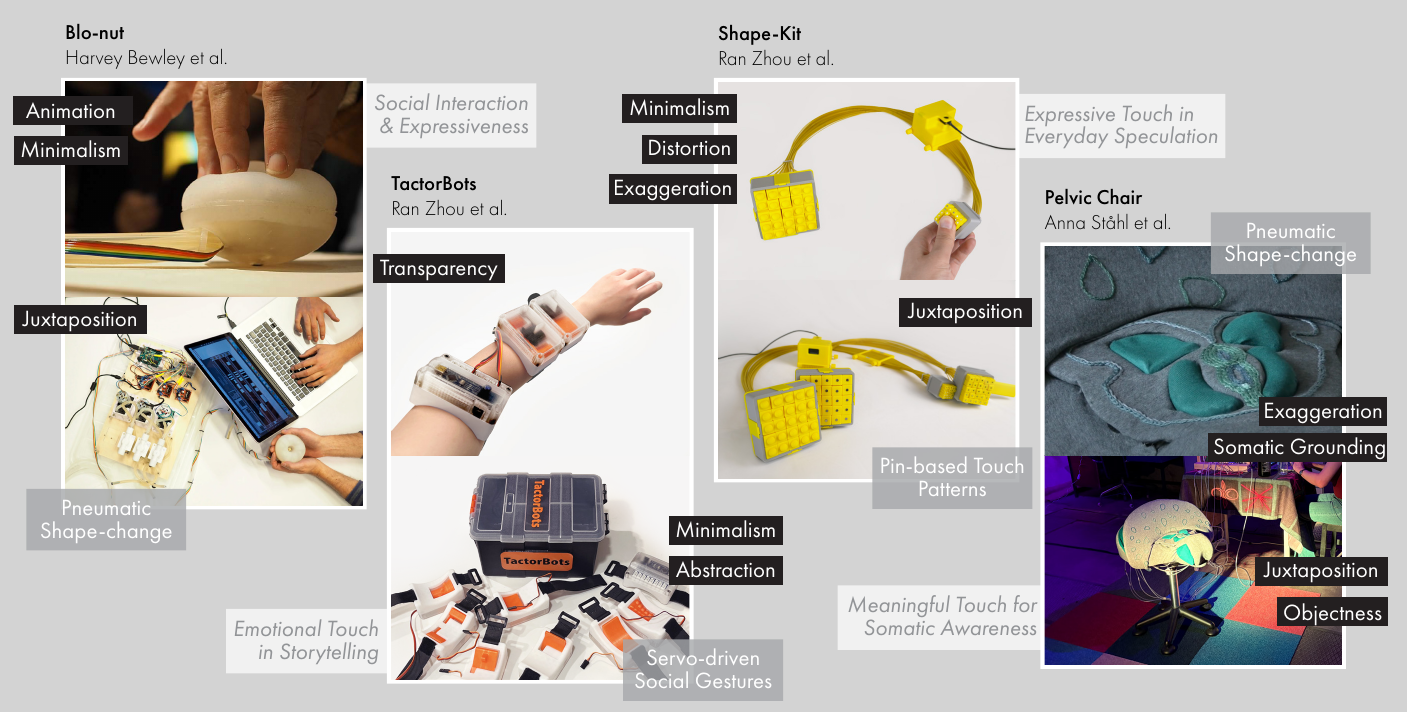}
  \caption{Four design research cases: Blo-nut \cite{bewley2018blo0nut, boer2018reconfiguring_blo-nut}, TactorBots \cite{zhou2023tactorbots}, Shape-Kit \cite{zhou2025shape-kit}, Pelvic Chair \cite{stahl2022annotated, yadav2025somatic_literacy, stahl_validity_21}. Annotation: robotic touch modality (dark gray background), expressive quality and context (white background), design strategy (black background)}
  \Description{Figure showing four robotic touch systems: Blo-nut, a soft donut-shaped robot with inflatable chambers for abstract expressive motion; TactorBots, small wearable modules that deliver simple tactile gestures; Shape-Kit, a pin-based toolkit that translates hand movements into dynamic tactile patterns for collaborative exploration; and Pelvic Chair, a soft sittable chair with pneumatic pockets that inflate to cultivate awareness of the pelvic floor. Each example is annotated with its modality, expressive qualities, and design strategies.}
  \label{fig:teaser}
\end{teaserfigure}



\maketitle


\section{Introduction}

Touch, our primary sense for physically exploring the world, is also a powerful medium for communication, \revise{emotional attunement, and aesthetic appreciation} \cite{bremner2012multisensory, classen2012deepest, hook2018designing}. Taking touch as a design material, we envision a future in which robotic touch is expressive and evocative, and capable of interacting with human bodies in meaningful and respectful ways. Within HCI, the touch channel has been mainly investigated by the haptics research community, with a primary focus on technology development. Haptic actuators are becoming increasingly miniaturized and compliant, capable of providing high-resolution, high-bandwidth stimulation \cite{yin2021wearable, ankit2022soft}. However, their explored touch has focused mainly on replicating real-life environmental cues for study- or demo-oriented applications in VR or gaming, where touch often serves as a secondary complement to visual or auditory interaction \cite{Culbertson_18_haptics, biswas2021haptic, frisoli2024wearable}. Despite being valuable, these directions leave much of touch’s everyday and expressive potential underexplored \cite{maclean2008haptic, jewitt2021manifesto, boer2017hedonic}.

Recent research in affective haptics and digital social touch has begun to investigate the communicative potential of technology-initiated touch, moving beyond \camera{simulating} how humans use their hands to explore static objects or surfaces \cite{huisman2017social, eid2016affective, mclaren2024affective}. This shift has expanded the horizon of what touch technology can attempt. But in lived experience, most active touch initiated by another subject comes from humans or animals \cite{loken2010skin, jones1985naturalistic, hertenstein_communication_2009, borgstedt2026uncannytouch}. Consequently, the replication of human touch behaviors, especially hand gestures, has become a new dominant paradigm \cite{mcintyre2021language, huisman_tasst_2013, simons2020incontact, haynes2019wearable, muthukumarana_touch_2020, stanley_design_2011}. Achieving high-fidelity replication, however, is not only technically challenging but also theoretically questionable, since interpersonal touch is a deeply embodied action. Researchers have started to question people’s willingness to replace interpersonal contact \cite{ryokai2022shifting, jewitt2021manifesto} and pointed to the risk of “social disfordance” in mediated touch \cite{mejia2017disfordance}. Moreover, the constant pursuit of realism in human touch replication risks slipping into the uncanny valley \cite{mori2012uncanny, berger2018uncanny}.

Looking at these tensions, we argue that the dilemma lies less in robotic touch itself than in the narrow replication viewpoint. \revise{In this paper, we aim to unpack and argue for an alternative stance that is gaining traction in HCI research.} Acknowledging the lack of a shared language for designing touch \cite{schneider2017haptic, jewitt2021manifesto}, we also recognize the challenges of inventing touch experiences that are relatable or registerable by humans without grounding them in something familiar. This is one reason why replication often serves as a natural starting point when exploring a new medium. Similar patterns can be found in art history, where realism dominated early practice before alternative movements such as expressionism and abstraction emerged to expand the space of creative possibility. We argue that now is the time for touch design to take such a step forward. 

As a highly interdisciplinary space, this provocation draws inspiration from adjacent communities. Most relevantly, in Human–Robot Interaction (HRI), researchers have argued for treating robots as an “other” subject \cite{sandry2015re}, or “otherware \cite{Hassenzahl_20}.” This stance has led to designs that move beyond dominant anthropomorphic or zoomorphic paradigms by staying true to the robot’s mechanistic nature, remaining their inherent “otherness.” Such an orientation not only avoids the pitfalls of over-expectation and the risk of uncanniness, but also opens a broader spectrum of interpretation and engagement, offering increased opportunities for developing long-term relationships with robots \cite{sandry2015re, Hassenzahl_20}. A similar sensibility resonates with design research, where Gaver et al. \cite{Gaver_03} argued for appreciating ambiguity as a resource. At a time when HCI was dominated by usefulness and usability, they showed how ambiguity could instead encourage meaningful engagement and foster more personal relationships with technology.

Inspired by these viewpoints, we propose treating robotic touch’s inherent \textbf{\textit{otherness}} as a design quality. Rather than a limitation in the pursuit of realistic replication, otherness can be embraced to elicit ambiguity that provokes alternative interpretations and creative imagination. Such a shift broadens the design opportunities for expressive robotic touch, opening ways for it to become more \textbf{\textit{acceptable}}, \textbf{\textit{intriguing}}, and \textbf{\textit{evocative}} \revise{over extended periods of time}. While informed by HRI theory, touch itself carries unique opportunities and constraints: it is always embodied, intimate, and immediate, while also deeply entangled with social norms and bodily boundaries \cite{price_2022_making_meaning, price2021conceptualising}. When contextualizing otherness in robotic touch, these factors are central to our inquiry. Our paper examines how they play out when robotic touch is designed with otherness, and how they may be reshaped by this intervening quality.




\revise{This paper makes a theoretical-analytical contribution to the design, haptics, and tangibles communities in HCI by arguing for otherness as a design quality for expressive robotic touch. Grounded in relevant HRI theories, we develop this argument through reflective Research through Design (RtD) approach \cite{Gaver_03, odom2016product, DiSalvo2014public} supported by the Annotated Portfolio Method \cite{gaver2012annotated}. We analyze inspirational art and design precedents alongside four design research cases with which we have intimate familiarity \cite{wensveen2014prototypes}. Through this analysis, we articulate a set of design languages for expressive robotic touch with otherness, comprising: \textbf{\textit{dimensions}} of touch meaning-making (\textit{\textbf{why}} otherness matters), \textbf{\textit{strategies}} for giving form to otherness (\textit{\textbf{how}} otherness is shaped), and \textbf{\textit{attributes}} of robotic touch system (\textit{\textbf{where}} otherness is embedded). These design languages can serve as a reflective resource \cite{lenz2013exploring} to support designers in thinking with and working through robotic touch design. We conclude by reflecting on tensions and risks in designing with otherness, highlighting how fidelity, transferability, and trust must be critically negotiated when bringing expressive robotic touch into everyday contexts.}

\section{Contextualizing Otherness}

In this section, we contextualize otherness in the design of expressive robotic touch. Similar to how ambiguity is more relevant in everyday contexts than in well-defined, safety-critical tasks \cite{Gaver_03}, the quality of otherness also has its contextual constraints. 
Many industrial robots, for example, are designed for precise control in the pursuit of efficiency. Likewise, certain haptic technologies for critical applications such as accessibility \cite{shull2015haptic}, medical treatment \cite{escobar2016review}, or safety training \cite{abate2009haptic} rely on accuracy and low-latency replication to ensure reliability. However, as robotic touch becomes increasingly present in everyday life, new opportunities arise for it to be not merely functional, but also expressive and evocative, \camera{similar to how ludic design supports activities that are motivated by curiosity, exploration, and reflection \cite{gaver2004ludic}. While such exploration may not be appropriate for all contexts at the beginning, probing these qualities can help us design better touch and, in the longer term, offer inspiration for how utilitarian systems might engage users beyond efficiency alone.}
In the following, we first trace how the notion of otherness has been discussed in HRI, then clarify our scope of otherness in the context of expressive robotic touch. 

\subsection{Theoretical Background of Otherness in Human-Robot Interaction}

The term “robot” was first proposed in Karel Capek’s play R.U.R (Rossum’s Universal Robots) in 1920 \cite{capek2004rur}, which was used to describe the creation of human-like replacements for people at work. Similar to many early fictions \cite{asimov2013robot, short2003measure}, the traditional strategy of designing robots has been to mimic humans or animals. This approach is based on the assumption that familiar appearances and modes of communication are the best way to support easy, effective, and accurate human-robot interactions \cite{sandry2015re, breazeal2004designing, breazeal_how_1999}. As a young interdisciplinary research field, HRI is still facing the challenges of establishing shared research methods and developing theories based on existing evidence. Findings from human behaviors and interpersonal communication, therefore, are commonly adopted as the theoretical basis when designing a novel interaction system \cite{hoffmann2017robot, Hassenzahl_20}.

At the same time, researchers have pointed out the limitations of such replication. While human-like robots are designed to reduce uncertainty and misunderstanding in the interaction, which might be helpful over the short term, several scholars argue that leaving a machine-like robot’s form and behavior open to alternative interpretations \cite{alves_21_collection, Hassenzahl_20, Anderson_2018_greeting} may offer increased opportunities for a long-term relationship to develop \cite{sandry2015re}. On the other hand, the concept of “uncanny valley” proposed by Masahiro Mori et al. \cite{mori2012uncanny} suggests that if a robot’s form and behavior are too similar to a human, but that it simultaneously is still clearly a robot, it will evoke a feeling of creepiness. Even perfect replicas of humans might be problematic because too much similarity blurs category boundaries that undermine human uniqueness \cite{ferrari2016blurring, hoffman2020social}. In response, HRI researchers have started to argue the value of non-anthropomorphic robots \cite{Anderson_2018_greeting, hoffman2015lamp, hu2024plant, koike2024sprout, walters2008avoiding}. From a design perspective, Hoffman et al. \cite{hoffman2014designing} identified benefits of building non-anthropomorphic robots, including freedom of exploration, economic feasibility and rapid prototyping, and the potential for higher acceptance.

Researchers \cite{sandry2015re} also proposed the strategy of tempered anthropomorphism and zoomorphism, where a robot’s behavior can be familiar enough to be interpreted as meaningful, while the existing otherness quality helps it leave space to acknowledge its fundamental differences from humans or animals. This is based on the fact that human understandings of robots, even machine-like robots, inevitably draw on anthropomorphic or zoomorphic responses with perceived animacy \cite{erel2019social, sandry2015re}. For instance, many Roomba users give it a name or consider it a family member \cite{alves_21_collection}. Rather than being a problem for otherness, HRI researchers consider it a valuable response \cite{sandry2015re}. Regarding those robots as “living” individuals could help build trust and relationships between them. As the robot is not designed in an explicit form familiar to humans, the relationship they build could be personal and unique. However, also because of its machine-like feature, people can always retain a clear acknowledgment of the otherness of the robot so that they would also be understood as machines that can be switched off and discarded.

Resonating with this perspective, Hassenzahl et al. \cite{Hassenzahl_20, sassmannshausen2023human} proposed the notion of “otherware” to move beyond seeing technologies as tools or as embodied extensions of humans. Instead, autonomous systems and robots can be envisioned as counterparts, where “technology becomes other.” They also discuss that designing robots by staying true to their mechanistic nature may also prove advantageous in social encounters, avoiding the pitfalls of over-expectation and opening a broader spectrum of possible responses. 

For social robots in design research, Boer and Bewley \cite{boer2018reconfiguring_blo-nut} argue that expressiveness need not rely on visually oriented sameness, which narrows the design space, but can instead emerge through explorations of form, material, and movement. By acknowledging robots’ otherness and centering on their dexterous movements, their appearance can privilege novel forms of expressivity. They emphasize that people’s tendency to anthropomorphize robots \cite{breazeal2003toward} arises primarily from their relation with them rather than from appearance alone. Embracing the otherness of social robots, they argue, can be valuable both for designing and for cultivating intriguing relations with them.

Overall, we see the term otherness in HRI as a call for the exploration of alternatives to realistic replication, or even the anxieties of replacement of humans or animals \cite{darling2021new}, by seeing robots as other subjects \cite{Diana2020hybridity} that retain their distinct qualities. Yet, while HRI and HCI research argue for highlighting the value of otherness beyond dominant anthropomorphic or zoomorphic paradigms, it does not prescribe what that “other” should be \cite{Hassenzahl_20}. Rather, it opens a space for alternative framings, interactions, and relationships.

\begin{figure*}[ht]
  \includegraphics[width=\textwidth]{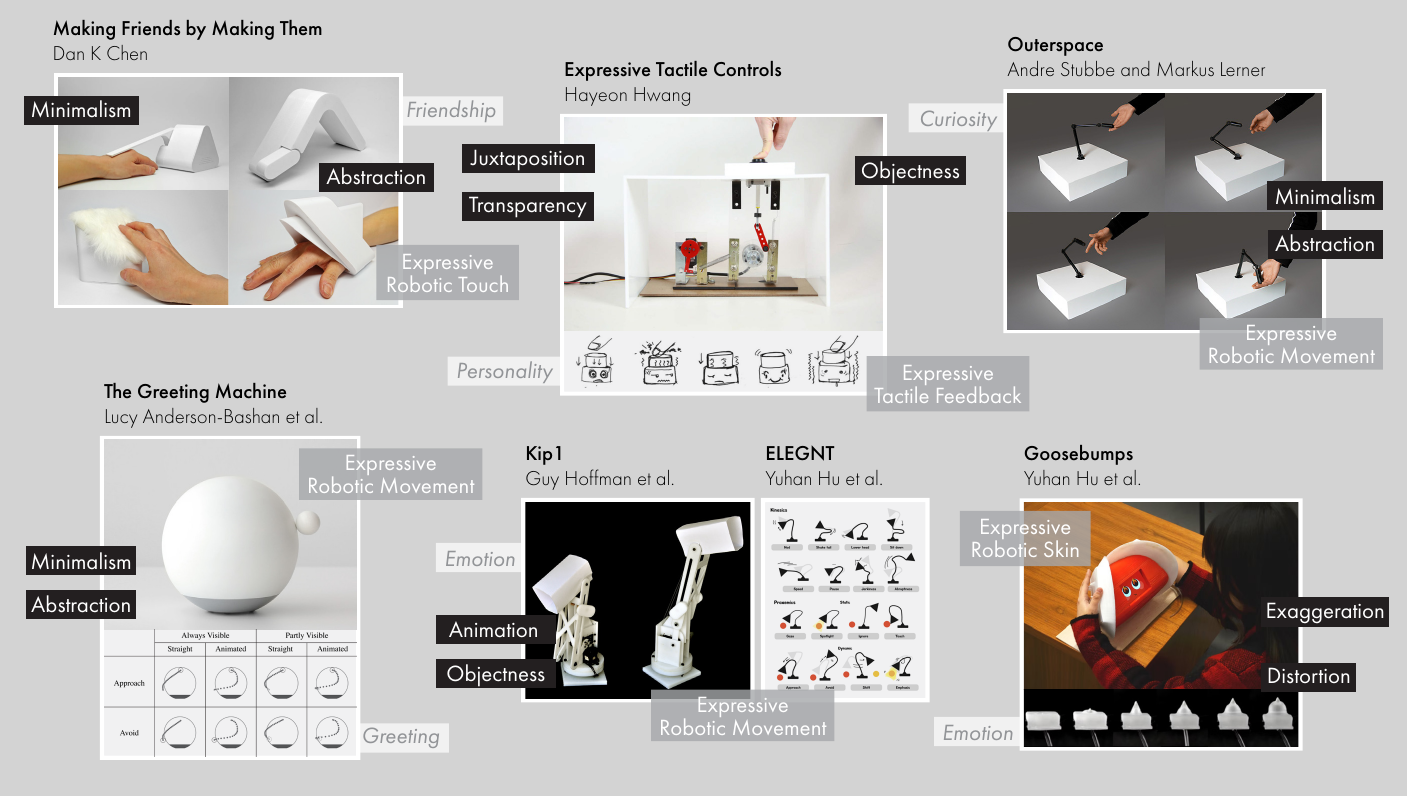}
    \caption{Annotated Precedents of Expressive Robotics: Making Friends by Making Them \cite{chen2021friends}, Expressive Tactile Controls \cite{hwang2019expressive}, Outerspace \cite{stubbe2006outerspace}, The Greeting Machine \cite{Anderson_2018_greeting}, Kip1 \cite{hoffman2015lamp}, ELEGNT \cite{hu2025elegnt}, Goosebumps \cite{hu_using_2019}. All images reproduced with permission. Copyright remains with the original authors/designers. Annotations: expressive modality (dark gray background), expressive quality (white background), design strategy (black background).}
  \Description{A collage of seven inspirational robot designs from art and HCI research. Top row: Making Friends by Making Them (small triangular robots showing minimal forms and companionship behaviors), The Greeting Machine (a globe-like platform with a moving ball signaling approach or avoidance), and Outerspace (a line-shaped interactive robot that responds playfully to touch). Bottom row: Expressive Tactile Controls (buttons that give different tactile personalities), Kip1 (a lamp-like robot showing curiosity through movement), ELEGNT (a similar lamp-like robot with expressive social behaviors), and Goosebumps (a robotic skin with spikes that inflate to show emotion). Each example is annotated with expressive modalities, qualities, and design strategies such as minimalism, abstraction, objectness, transparency, exaggeration, distortion, animation, and juxtaposition.}
  \label{fig:inspirational_works}
\end{figure*}

\subsection{Otherness of Robotic Touch}

Stemming from the notion of otherness in HRI, \revise{we argue that it can be productively adapted to robotic touch, with implications for the broader haptics and tangible interaction communities in HCI.} Resonating with HRI debates, we acknowledge the challenges and limitations of the ongoing pursuit of realistic replication \cite{ryokai2022shifting, jewitt2021manifesto, mejia2017disfordance, berger2018uncanny}. However, rather than aiming to criticize or replace well-established research on programming humanoid robots to touch humans \cite{mazursky2022physical, hoffmann2017robot}, uncovering the scientific knowledge of touch perception \cite{hauser_uncovering_2019, mcintyre2021language, xu_3d_22, loken2009coding}, or creating natural haptic sensations that are no longer intrusive or clunky \cite{shen2023fluidreality, ankit2022soft}, our focus is to intervene in these mainstream trajectories and explore alternatives. We seek to expand the design space by asking what else robotic touch can do, and how it might become more acceptable, intriguing, and evocative, especially in ways that resonate with creativity-oriented communities such as art and design. 

Otherness arises when robotic interfaces serve as “other” subjects and acknowledge their \camera{mechanistic }nature, and as such do not strive for anthropomorphic or zoomorphic replication alone. In the context of touch, this can also be specified to not replicate the felt sensation of being touched by a human or animal. Otherness is not new to robotic touch. Many low-fidelity systems that attempt replication \cite{mcintyre2021language, huisman_tasst_2013, simons2020incontact,muthukumarana_touch_2020, stanley_design_2011} inevitably reveal their machine-like qualities through jitter, constrained movement, or mismatched modalities, which inherently sustain their otherness. Yet these are typically framed as technical shortcomings, since the research goal has often been to test whether sensations can be recognized as realistic environmental cues or gestures. In other cases, especially in high-performance on-skin interfaces, actuators are developed to deliver accurate, recognizable gestures \cite{haynes2019wearable, Kim_KnitDermis_21, hamdan2019springlets}. While technically impressive with great potential in everyday haptics, the focus remains narrow, where success is framed as faithful replication of known gestures. We pose the question of what design possibilities open up beyond that step. In line with tempered anthropomorphism \cite{sandry2015re}, we see familiar touch, such as those gestures, as useful references for identifying different types of tactile sensation, but accurate replication should not be the ultimate goal. Robotic touch can and should engage with broader design possibilities, as it need not be human-like to be expressive and meaningful.

We therefore argue for treating robotic touch’s inherent otherness as a design quality. Instead of being a limitation for realism, otherness can be embraced, appreciated, and strategically designed with to elicit ambiguity, provoke alternative interpretations, and expand the design space of expressive touch beyond stereotypes of living beings’ touch behavior mimicry. Following HRI, where otherness has been articulated but not strictly defined, we do not seek to prescribe what otherness “is” in robotic touch once and for all. Instead, we treat otherness as an evolving concept, entangled with robotic system design, bodily experience, and social context. Rather than \camera{establishing a fixed definition or set of design guidelines}, our aim is to develop languages that help designers think with and work through otherness in practice.

\section{Otherness Shaped in Art \camera{\&} Design Precedents}
\label{subsec:art_design}
To explore how such otherness can be given form, we turn to the art and design realm, where non-anthropomorphic robotic interfaces have been explored for expressive, engaging, and aesthetically compelling interactions. Following the Annotated Portfolio Method \cite{gaver2012annotated, hoggenmuller2021elciting, gaver2012expect}, we curated a set of inspirational precedents (Fig. \ref{fig:inspirational_works}). Our selection reflects the design inspirations that have shaped our own exploration: while we have not experienced all of these works in person, we studied them through videos, articles, and publications, and in some cases, direct conversations with the designers, which provided insight into their motivations and process. This collection is also chosen to illustrate a wide variety of expressive modalities and design approaches. The first row presents art and design installations that have been exhibited publicly, while the second row includes robots designed within HRI research. We do not intend this as an exhaustive collection, but rather as a way to set the tone for our argument: these examples illustrate how robotic interfaces can be highly expressive while maintaining a sense of their “otherness.” We annotated the precedents by their expressive modality and quality. \revise{From this proces we also identified and annotated design strategies that are recurring or singular but striking, which we see can inspire how otherness might be shaped:}

\textbf{\textit{Abstraction \& Minimalism.}} These strategies resonate with the “form follows function” principle in architecture and industrial design. Using basic geometric forms and limiting degrees of freedom (DoF), robotic interfaces can embody expressive behaviors without anthropomorphic detail. Outerspace \cite{stubbe2006outerspace} takes a line-like form, conveying curiosity through reactive movements that invite touch and play. The Greeting Machine \cite{Anderson_2018_greeting} features a small ball moving across a globe to signal approach or avoidance in initiating engagement. Making Friends by Making Them \cite{chen2021friends} is a series of triangular-prism–shaped robots that offer simple touch behaviors as companionship. These examples show how minimal forms allow the design to focus on the delicacy of nuanced interactive elements, such as movement, path, and touch. It can sustain salient expressiveness while maintaining high aesthetic quality for appreciation and gestalt delight in interpretation.

\textbf{\textit{Objectness \& Animation.}} These two strategies can work together to preserve the recognizable “objectness” of everyday artifacts while animating them through robotic mechanisms to evoke expressive interpretations. Such objects are non-living by nature, yet deeply integrated into daily life through their cultural affordances, resonating with Dunne and Raby’s technological dreams series \cite{dunne2024speculative} \revise{and Objects with Intent \cite{Rozendaal2019OwI}}. Expressive Tactile Controls \cite{hwang2019expressive} take the mundane push button and endow it with personality (e.g., timid, stubborn, impatient) through distinct tactile feedback when pressed. Similarly, lamp-like robots such as Kip1 \cite{hoffman2015lamp} and ELEGNT \cite{hu2025elegnt} retain the familiar presence of a domestic desk lamp, yet through carefully designed kinesthetic and proxemic movements, they convey curiosity and social stance. In these examples, otherness emerges from the tension between the static familiarity of the object and its newfound animated agency.

\textbf{\textit{Exaggeration \& Distortion.}} These strategies are inspired by abstract art techniques \cite{schapiro1937nature}. They help unfamiliarize the familiar behavior or the phenomena of living beings into robotic expression by extracting their qualities, exaggerating their dynamics, and distorting their scale. Goosebumps \cite{hu_using_2019} draws inspiration from how humans and animals signal internal states through skin texture changes (e.g., human gooseflesh, a cat’s raised fur, or a blowfish’s spikes). The project re-imagines these micro-expressions as robotic skin shape-change, exaggerating their motions and distorting their shape and scale. By extracting \camera{and reconfiguring} these features into forms compatible with robotic mechanisms, the project creates expressions that are at once evocative of the familiar but distinctly other qualities. While the robot in Fig. \ref{fig:inspirational_works} includes animated eyes to illustrate a possible use case, the research focused specifically on the expressive potential of the shape-changing skin itself.

\textbf{\textit{Juxtaposition \& Transparency.}} This pair is to juxtapose the simplicity of form on the surface with the transparency of its underlying complexity. Expressive Tactile Controls \cite{hwang2019expressive} exemplify this: while the push button is among the simplest controls, its expressive tactile feedback is generated by sophisticated robotic mechanisms that are deliberately made visible to the audience. By exposing the hidden process beneath the surface, the work enhances the robot’s perceived honesty and foregrounds its otherness \cite{hu_what_22}. The juxtaposition between the minimal 1 DoF motion and the intricate actuation system that enables expressive feedback not only generates surprise but also invites deeper sense-making. 

While designers and artists often make decisions guided by their own sensibilities, across these precedents, we see that the shaping of otherness is not arbitrary. \revise{The strategies we identify are not presented as an exhaustive set, but as generative handles to} help trace these design decisions, showing how they contribute to systems with relatable expressiveness while leaving room for ambiguity. We introduced the strategies in pairs here \revise{to highlight their complementary relationships}, but as the later case studies show, they can also be applied individually or intertwined. 

Among the art and design precedents, Dan K. Chen’s Robot Friends \cite{chen2021friends} are particularly aligned with our theme on robotic touch. In his \revise{documentation} \cite{Chen2012file}, Chen reflected on moving away from human skin replication with latex, which was perceived as uncanny in his early experiments, toward abstract, minimal forms: “The design of my robots is honest with its function. Using no fancy adornments, I do not attempt to disguise the robots or portray them as anything but what they are.” We see this as a strong articulation of embracing, respecting, and designing with robotic interfaces’ inherent otherness.
\begin{figure*}[ht]
  \includegraphics[width=\textwidth]{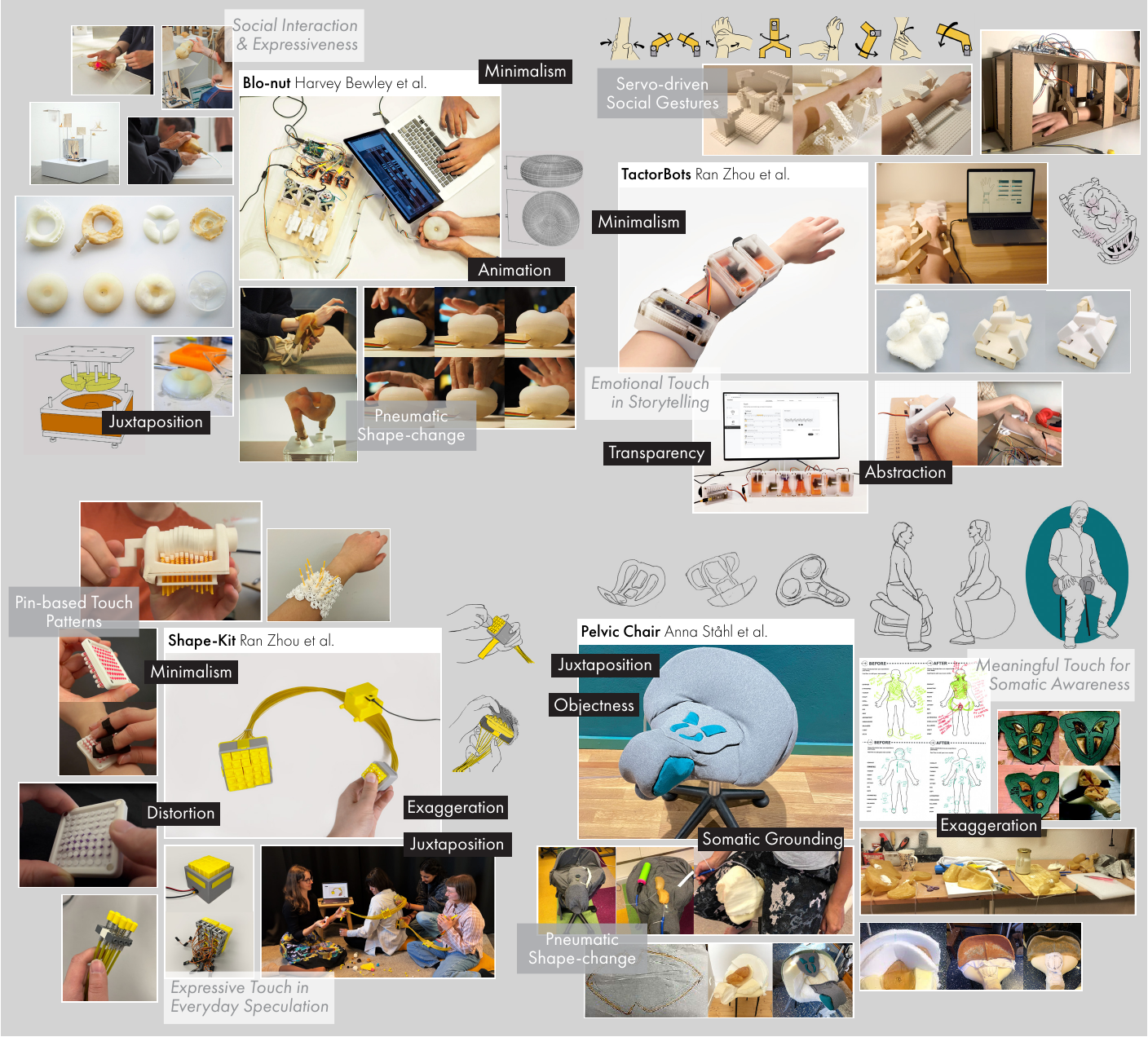}

    \caption{Four design research cases with prototyping \& iteration collages: Blo-nut \cite{bewley2018blo0nut, boer2018reconfiguring_blo-nut}, TactorBots \cite{zhou2023tactorbots, zhou_emotitactor_2020, zhou_emotitactor_22}, Shape-Kit \cite{zhou2025shape-kit}, Pelvic Chair \cite{stahl2022annotated, yadav2025somatic_literacy, stahl_validity_21}. Annotation: robotic touch modality (dark gray background), expressive quality and context (white background), design strategy (black background)}
  \Description{A large collage-style figure composed of four clustered design research cases of expressive robotic touch, each surrounded by images documenting prototyping and iterative exploration. The Blo-nut section (top left) shows soft, donut-shaped pneumatic actuators in various forms and deformation states, alongside images of hands pressing and shaping them, a lab setup with electronics and a laptop, and sequential frames illustrating animated shape changes. Annotations highlight qualities such as social interaction and expression, pneumatic shape-change as modality, design strategies are: animation, minimalism, and juxtaposition. The TactorBots section (top right) presents a wearable forearm device composed of modular tactile units. Surrounding images include servo-driven mechanisms, small sculptural forms, on-body interaction scenarios, and interface screenshots. Annotations emphasize servo-driven social gestures, emotional touch in storytelling, with strategies of abstraction, transparency, and minimalism. The Shape-Kit section (bottom left) features a modular pin-based haptic toolkit. Images show handheld devices, pin arrays producing tactile patterns, exaggerated and distorted forms, and a workshop setting where participants collaboratively explore touch designs. Annotations include pin-based touch patterns, expressive touch in everyday speculation, with strategies of minimalism distortion, exaggeration, juxtaposition. The Pelvic Chair section (bottom right) depicts a soft, chair-like object with a blue cushioned form mounted on a base. Surrounding images include sketches, foam prototypes, molding processes, and body maps. Annotations highlight pneumatic shape-change, meaningful touch for somatic awareness, with strategies of objectness, somatic grounding, exaggeration, and juxtaposition.}
  \label{fig:iteration}
\end{figure*}

\section{Design Research Cases}
\label{robotic_touch_cases}
In this section, we describe four design research projects \camera{(Fig. \ref{fig:teaser}\&\ref{fig:iteration})}, which have begun to explore otherness as a quality, either explicitly or implicitly, in the design of expressive robotic touch. We briefly introduce each case by highlighting their motivations, system designs, and empirical explorations, and how they engaged with otherness as well as some backstories shared by their designers. A more analytical account of where and how otherness is embedded, and why it matters for meaning-making, is presented in Sections \ref{design language}.

\textbf{\textit{Selection Criteria.}}
We deliberately selected cases with which we have an intimate familiarity \cite{wensveen2014prototypes}\camera{, as we and our collaborators are the original designers.} This allows us to deepen our analysis to surface design rationales and experiential nuances that we could not include otherwise. 
The cases include one provocative artistic and research artifact, Blo-nut \cite{bewley2018blo0nut, boer2018reconfiguring_blo-nut}, originally intended to explore visual expressive motion of the soft robot, but in which touch emerged as inevitably significant, and two design toolkits, TactorBots \cite{zhou2023tactorbots} and Shape-Kit \cite{zhou2025shape-kit}, which enable in-the-wild and collaborative explorations of expressive touch, respectively. To our knowledge, these three projects are the only works that explicitly discuss otherness in the context of expressive robotic touch within the closest communities (ACM CHI and DIS). We also include the Pelvic Chair \cite{stahl_validity_21, stahl2022annotated, yadav2025somatic_literacy}, a research product designed to foster awareness and somatic literacy of the pelvic floor through robotic intimate touch \cite{balaam2020intimate}. Although this work does not use the term otherness directly, it has a strong claim of resisting mimicking human touch. We consider it a valuable example of how robotic touch can be carefully designed to provide meaningful interactions on intimate body parts, with otherness implicitly embedded in the design. 


\subsection{Blo-nut}
Blo-nut \cite{bewley2018blo0nut, boer2018reconfiguring_blo-nut} is a provocative artistic object and research artifact \camera{(Fig. \ref{fig:teaser}\&\ref{fig:iteration})}. Motivated by their formative explorations in Performing Objects \cite{bewley2017performing}, Blo-nut was created to challenge the assumption that social robots should be visually human- or animal-like, instead advocating for an explicit acknowledgment of their otherness. The authors describe otherness as “an aesthetic category broadly used to describe robotic forms that are neither anthropomorphic nor zoomorphic in their outward form.” In this spirit, Blo-nut was designed as a deliberately abstract form: a smooth silicone donut with three inflatable chambers mounted on a rigid plastic base. Its hybrid construction played on the contrast between an engineered aesthetic and the elastic soft-robotic movement.

To probe its expressive potential, three experts were invited to engage in “choreographic sketching” with Blo-nut using a music-inspired graphical user interface (GUI). A plurality of associations emerged during their design process, from  \textit{“sea creatures”} and \textit{“frog ponds”} to \textit{“alien rituals”} and \textit{“mating dances,”} underscoring the interpretive openness of Blo-nut’s abstract form. Participants further emphasized its multi-sensory presence: the sound of pumps and valves, together with the tactility of the inflating silicone skin, became inseparable from its expression. As one expert described, the object felt \textit{“very not-human, nor animal, but it still has life.”}

Blo-nut was staged in a fictional future scenario, as the first prototype of a possible “otherness pet” to be evaluated outside the labs in which they were developed. By performing the expressive choreographies created by the experts, three groups of designers interacted with it to imagine the emerging robot relationships. Their impressions ranged between seeing it as alive yet mechanical, pet-like yet product-like. These ambivalent readings offered no straightforward conceptualization but instead pointed to the curious and provocative relationships that can emerge. More broadly, the work argues that acknowledging otherness in the process of giving form to social robots, and in growing the relationship with them, can open space for alternative forms and interactions.

\subsection{TactorBots}
TactorBots \cite{zhou2023tactorbots} is a haptic toolkit with wearable robot modules for exploring emotional touch. It grew out of the authors’ earlier work with EmotiTactor \cite{zhou_emotitactor_2020, zhou_emotitactor_22}, which began by recreating social gestures inspired by interpersonal emotional touch but later uncovered alternative design metaphors beyond human hand behaviors \camera{(Fig. \ref{fig:teaser}\&\ref{fig:iteration})}.  For instance, one designer believed a small tactor shaking slowly and gently beneath the wrist could express Love, since it reminded them of the sway of a warm cradle. Insights like these indicated the potential of otherness in abstract robotic tactors. TactorBots was designed to explore such alternative opportunities more intentionally.

The toolkit consists of eight small modules that can be strapped to the body and can be controlled by a web GUI. Each module was designed to render a distinct touch gesture such as pat, squeeze, or rub. Rather than resembling hands in form or texture, the modules were deliberately machine-like, fabricated in plastic, each driven by a single micro-servo, with movements made visible through a clear enclosure. The authors describe their approach as “embracing ambiguity in emotional haptic design by leveraging the balance between the familiarity of interaction—recognizable social gestures—and the otherness of form—the robotic tactor’s appearance and texture.” 

In a long-term deployment, thirteen designers used TactorBots to prototype emotional touch for enriching a fictional story. While the tactors were drawn on social gestures, many designers focused on each module's raw perception, treating them as open-ended materials for design. For instance, the Rub tactor with its spinning barrel of raised dots, was described as evoking associations like \textit{“goosebumps”} or the \textit{“spines of a cat’s tongue.”} When designing the touch, most designers did not picture human-human interaction as inspiration. Instead, they could tactilize some metaphors like \textit{"steam coming out of ears"} to convey Anger or \textit{"surrounded by dancing butterflies"} for Happiness. Even if using human or animal metaphors, inspiration often came from bodily responses like \textit{“choking”} or animal behaviors such as \textit{“cat’s licking,”} or \textit{“dog’s barking”} rather than imagining a hand or paw. Designers also explored body placements beyond the default socially acceptable forearm \cite{suvilehto2015topography}, attaching modules to areas such as the neck or ankle to test out various possibilities. Together, TactorBots illustrate how otherness in robotic touch can foster alternative perceptions, imaginative interpretations, and playful bodily exploration that move beyond the social norms typically associated with interpersonal touch.

\subsection{Shape-Kit}

Shape-Kit \cite{zhou2025shape-kit} is a hybrid design toolkit for collaborative “crafting haptics,” where designers can prototype diverse haptic experiences intuitively by touching with their hands \camera{(Fig. \ref{fig:teaser}\&\ref{fig:iteration})}. It consists of an analog tool that translates human hand manipulations into amplified or minified dynamic pin-based sensations through a flexible transducer, allowing free-form exploration of touch across the body or on another person’s body. An ad-hoc tracking module captures and digitizes the patterns, while a GUI supports recording and playback. Shape-Kit engages with otherness quality by intentionally lowering the resolution of hand touch through pixelation and distortion in scale, producing the outcome of robotic touches that feel rhythmically natural yet mechanically other.

Six teams of designers used Shape-Kit to explore expressive haptics for speculative everyday scenarios such as sleep \& wake-up enhancement and haptic white noise. Their exploration revealed the richness of touch design achieved through diverse crafting methods, from bare-hand techniques to the use of props and materials. Participants, for instance, crafted continuous waves across the chest when lying down to lull sleep, which they named “Sailboat Dreams”; or designed a narrative pattern to simulate the feeling of walking on sand, \textit{“your toes are sinking into it, water is rushing it like ripples”} to evoke a sense of nature indoors. 

Notably, Shape-Kit fostered collaborative design, which is often challenging in the haptic modality, given the intimacy of touch. With Shape-Kit, designers were able to ideate, prototype, iterate, and compose bodily haptic experiences together through rapid role-switching between crafter, perceiver, and holder. They also explored a wide range of body locations during the study (e.g., arms, shoulders, thighs, foot soles, back, chest, belly), extending beyond typical social areas \cite{suvilehto2015topography}. The authors reflected that this may have been enabled by the design of the long transducer, which maintained appropriate proxemics between crafter and receiver, and by the pin-filtered sensations, whose otherness neutralized the human touch, enhancing openness to bodily exploration that was playful, curious, and respectful.

\subsection{Pelvic Chair}

The Pelvic Chair \cite{stahl2022annotated, yadav2025somatic_literacy, stahl_validity_21} is a research product for raising awareness and somatic literacy of the pelvic floor through intimate robotic touch \camera{(Fig. \ref{fig:teaser}\&\ref{fig:iteration})}. Positioned as a critical feminist design intervention, it seeks to draw attention to the anatomy, muscles, and connectedness of this often neglected region, exploring touch as a way of knowing one’s own intimate body. The chair is made from soft materials, including foams and fabrics, embedded with latex inflatable pockets. As the authors describe: “When someone sits on it, parts of the chair inflate and move—to encourage the user to sit upright, to open their legs slightly. Then the inflatables touch different parts of the pelvic-floor muscles” \cite{balaam2020intimate}.

The development began from first-person soma design practice \cite{hook2018designing}, where the authors took Feldenkrais sessions to sensitize their own bodies and better understand the pelvic floor through estranging practices and subtle bodily movements. From this process, they identified experiential qualities—differentiation, unitedness, connectedness, and relaxational power—that were evocative and potentially meaningful over the long term. Pelvic Chair was designed to embody these qualities and deliver them directly to the pelvic floor muscles and skeleton through shape-change. With a focus on bodily awareness, the design deliberately avoided simulating existing cues or evoking associations with medical devices or intrusive hand-like touch—an approach that resonates strongly with designing for otherness. Instead, the inflatable pouches’ shape and placement were decided based on the organic shape of the body and the anatomy of the pelvic floor. When actuated, they expand into areas of less pressure so that certain regions feel softly held and gently raised.

Studies with participants showed that such touch could provide an active way of sensing the pelvic floor and cultivating somatic literacy. Despite engaging a typically taboo area, participants did not find the touch uncomfortable, sexual, or uneasy, but instead supple, intimate, and gentle. Many emphasized that the experience felt \textit{“not machine-like, not human-like,”} and maybe too novel to be tied to familiar metaphors or personal histories. Rather, it was appreciated as aesthetic and memorable, speaking directly to the body itself. With a strong sense of agency and the improved awareness of pelvic-floor muscles, several participants experimented with moving their body along the touch, while even doing finer control of their muscles to have a more united and playful engagement with the chair. Moreover, this work critically explores how robotic touch might approach intimate body regions in ways that are acceptable, safe, and comfortable, while also fostering meaningful and dignified engagement with the body.

\section{Design Languages for Expressive Robotic Touch}
\label{design language}


\revise{To extend our theoretical discussion of otherness into the context of robotic touch, we take reflective analysis across the four design research cases to examine \textit{\textbf{where}} otherness can be embedded within robotic interfaces and \textbf{\textit{how}} it is shaped across different design attributes. We then unpack \textbf{\textit{why}} otherness matters by analyzing its impact on touch \revise{meaning-making} across different dimensions, showing how it can contribute to robot touch that is acceptable, registerable, and expressive. Finally, we synthesize these insights with an overview of the design languages.}
 

\revise{\textbf{\textit{Analysis Approach.}}
Our analysis follows a reflective RtD approach \cite{Gaver_03, odom2016product, DiSalvo2014public}, supported by the Annotated Portfolio Method \cite{gaver2012annotated}. Rather than re-analyzing the individual cases through new empirical studies or post hoc coding, we approach them as design research artifacts embedded in long-term research trajectories, where knowledge emerges through making, use, and reflection over time \cite{redstrom2017making}. The analysis was conducted through iterative reflective discussions \camera{among us, as the original designers and researchers of the systems}, all of whom have sustained engagement, often over five years, in designing and thinking with otherness in robotic touch. These discussions drew on \camera{our} shared design histories, including early prototypes, iterations, pilot studies, design experiments, and evolving motivations \camera{(Fig. \ref{fig:iteration})}. This reflective synthesis enables us to identify patterns across cases, providing access to tacit design judgments and experiential insights that are difficult to uncover through individual short-term empirical studies or workshops.}

\begin{figure*}[ht]
  \includegraphics[width=\textwidth]{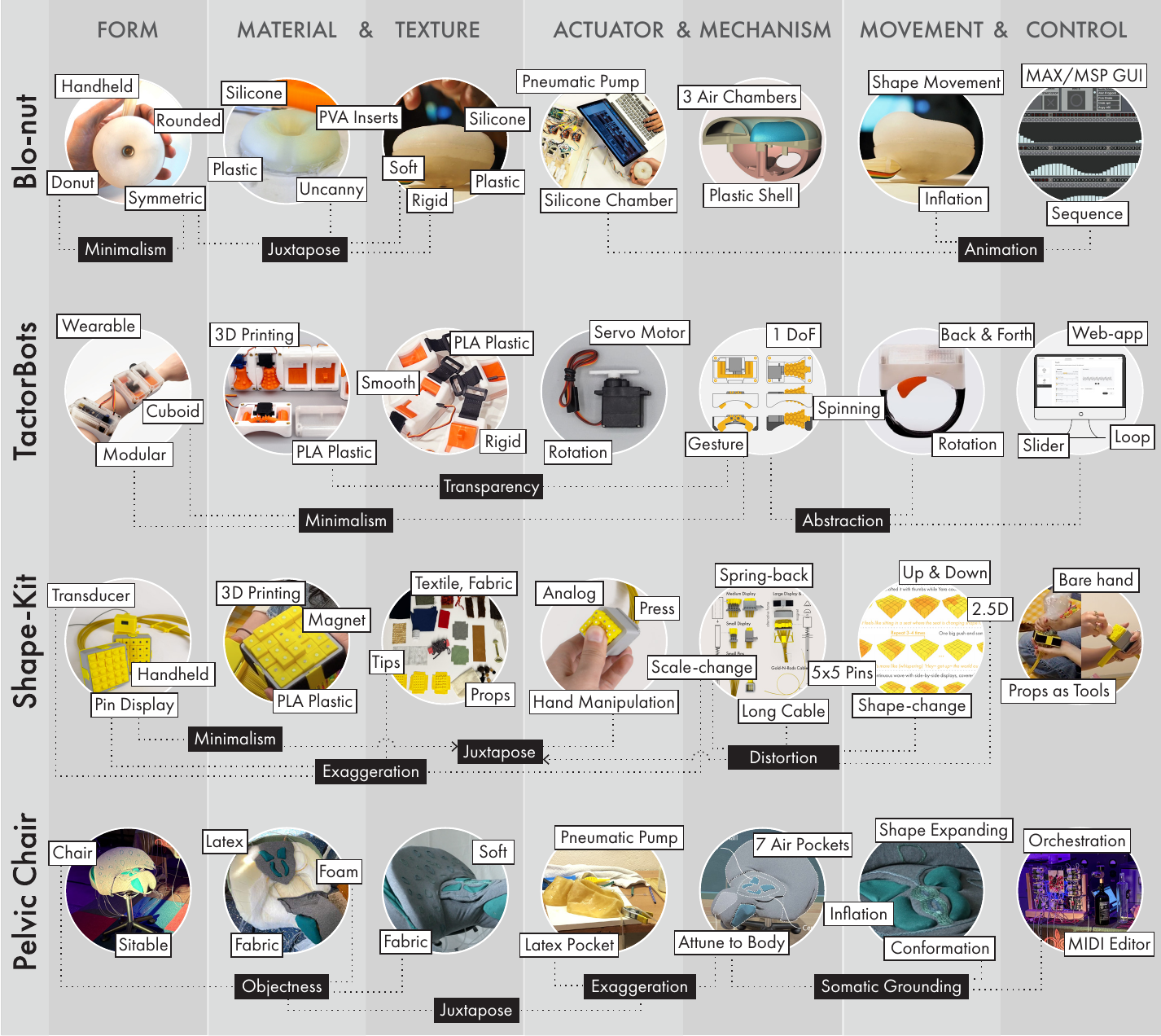}
 
    \caption{Deconstruction of four robotic touch systems into design and control attributes, with annotations highlighting attribute features (white background) and their connections to strategies shaping otherness (black background).}
  \Description{Figure comparing four robotic touch systems by deconstructing their design attributes. Categories include form, material, texture, actuator, mechanism, movement, and control. Examples show Blo-nut with a soft donut form and pneumatic chambers; TactorBots with modular PLA plastic enclosures and servo-driven gestures; Shape-Kit with pin-based displays actuated by hand manipulation; and Pelvic Chair with a sittable form using latex pneumatic pockets. Each system is annotated with features and linked to strategies such as minimalism, abstraction, transparency, juxtaposition, animation, exaggeration, distortion, and somatic grounding.}
  \label{fig:deconstruction}
\end{figure*}

\subsection{Where and How Otherness is Embedded}
\label{deconstruction}

Our case studies span provocative artifacts, design toolkits, and research products, combining rigid plastic modules and displays with soft, organic, or textile textures, and taking form factors such as hand-held devices, wearables, and sittable artifacts. Together, these cases demonstrate the variety of potentials in which otherness can emerge and be designed in robotic touch. We recognize that expressive touch is always a holistic experience, shaped not only by the hardware system and computational control, but also by environment, context, timing, and the receiver’s personality, personal history, and imagination \cite{Karpashevich2026monstrous}. \revise{As widely discussed in HRI, additional factors such as the social relationship between the robotic interface and its user, and questions of agency and consent regarding who initiates the touch, when, and where, can further influence how touch is perceived and interpreted \cite{hoffmann2017robot}. Resonating with Gaver et al.’s focus on how artifacts can be designed to elicit different forms of ambiguity \cite{Gaver_03}, we narrow attention here to the design of the robotic touch systems themselves, as this is the aspect most directly and intentionally shaped through design decisions.} Specifically, we zoom into system design attributes to analyze where otherness is embedded and how design decisions embody strategies of otherness, making them tangible in robotic touch.


\subsubsection{Layering Otherness in Robotic Touch Systems}

While otherness in HRI is often identified primarily in relation to a robot’s outward form \cite{bewley2018blo0nut, sandry2015re}, we argue that in robotic touch, otherness can be embedded across multiple design attributes that shape how sensations are rendered and experienced. Drawing on perspectives from haptic \revise{and touch interface design \cite{hannaford2016haptics}, we identified key attributes that structure the design of robotic touch systems. Although each attribute represents a distinct design concern, we pair up closely related attributes for analysis: \textit{form, material and texture, actuator and mechanism, and movement and control}.} \camera{As shown in Figure \ref{fig:iteration}, the design of these cases was not linear; rather, each has undergone a series of prototyping and iterations across attributes. However, as this paper focuses on high-level reflection, we primarily examine each system's final design decisions (Fig. \ref{fig:deconstruction}),} while referring readers to the original publications for full accounts \cite{bewley2018blo0nut, zhou_emotitactor_2020, zhou_emotitactor_22, zhou2025shape-kit, stahl2022annotated, stahl_validity_21}.  Figure \ref{fig:deconstruction} illustrates how we deconstruct the four case systems across attributes, and in the following, we explain where and how otherness is embedded for the four cases. 

\textbf{\textit{Form.}} Form in robotic touch interfaces includes form factor, scale, shape, and outward appearance. Unlike many haptic prototypes or lab workbenches that focus purely on tactile sensations, often by blindfolding participants or concealing the device within a VR scene, we argue that expressive robotic touch interfaces should be engaged in their full presence. In such contexts, form is crucial. It sets the audience’s first impression and conditions whether and how they are willing to interact. Prior replication-driven works have built robots that touch humans by mimicking human or animal bodies \cite{chen2011touched, yu_use_2015} or the form of hands or paws \cite{nakanishi2014handshaking, teyssier2018mobilimb, jiang2025miau-bot}. By contrast, we see the otherness emerges through reducing or avoiding extra details that would reinforce such an association, presenting designs that are distinctly robotic in their own right.

Across our cases, each system is carefully designed and executed while intentionally avoiding obvious aesthetic links to living beings or their touch organs. Blo-nut, TactorBots, and Shape-Kit keep the \textbf{\textit{minimalism}} in the form design, which reduces form to essentials and maintains an engineered, industrial design style with regular geometric shapes. Pelvic Chair instead draws on \textbf{\textit{objectness}}, beginning from a stool-like form—a familiar but non-living artifact—considered well-suited to let the pelvic floor rest and be touched while allowing the sitter to remain relaxed, comfortable, and feel safe. A front piece was added to reduce the sense of vulnerability when sitting with the legs slightly open, similar to sitting on a horse saddle. Across each design, the scale and shape give sufficient affordances to enable different modes of engagement (e.g., wearable, hand-held, or sittable), while many highlight actuation parts with different colors to make robotic behaviors legible. 

\textbf{\textit{Material \& Texture.}} These attributes encompass both the materials used to fabricate the system and the surface properties that come into direct contact with the body. Texture shapes the initial skin encounter, whether smooth or rough, sticky or slippery, fluffy or firm, while material governs ongoing engagement, influencing whether touch feels rigid or soft, stable or fluid. The surface material’s thermal conductivity also decides whether it feels cold or warm. Human perception of material and texture is rich, nuanced, and individually varied \cite{pasqualotto2020tactile}. For instance, softness may feel pleasant, but if paired with coldness or stickiness, it can quickly become disgusting or uncanny. These attributes thus open a broad space for exploration, including human affective associations with materials \cite{drewing2018material}. At the same time, certain materials, such as silicone \cite{teyssier2019skin-on}, latex \cite{bewley2017performing}, or long fur fabric \cite{zhou_emotitactor_22}, are strongly tied to anthropomorphic or zoomorphic interpretations. By contrast, our cases show how these associations can be negotiated in designing for otherness: sometimes deliberately acknowledged for juxtaposition, sometimes masked, and sometimes avoided by turning to neutral materials and textures.

Blo-nut combines a rigid plastic base with a silicone upper skin and PVA inserts, following their design principles of \textbf{\textit{juxtaposition}}: the intentional uncanniness of its bio-skin-like texture contrasts with its minimalist form. TactorBots and Shape-Kit employ PLA plastic, the most common FDM 3D printing material, which is rigid, lightweight, and neutral. While TactorBots align well with machine-like otherness, participant feedback revealed that plastic or metal shells may also fall into a stereotype of robots as “other.” Accordingly, their later Shape-Kit introduces attachable props from everyday fabrics, meshes, and custom tips that extend the texture quality of the touch, while the pins’ material remains firm. Here, texture becomes an explicit site of exploration, capable of \textit{\textbf{exaggerating}} touch with richer variations in feel. The Pelvic Chair relies entirely on soft materials to ensure comfort in intimate contexts. Its inflatable latex pockets are surrounded by foam, and both are wrapped in fabric, aligning its tactility with ordinary furniture and sustaining its \textit{\textbf{objectness}}.

\textbf{\textit{Actuator \& Mechanism.}} Actuators are the core elements that enable a robotic interface to touch the human body, while mechanisms decide how the robotic touch is performed. Whereas technical haptics research typically evaluates actuators in terms of torque, bandwidth, displacement, block force, or response rate, we approach actuators as design components. Similar to how HCI has come to value the materiality of raw materials in interaction design \cite{jung2011materiality}, we suggest that robotic touch requires attention to the felt qualities of actuation, and how the distinctive properties of different actuators can be creatively harnessed, together with attuned mechanisms, to produce expressive touch experiences. Across our cases, the systems primarily employed servo motors and pneumatics, perhaps the most approachable actuators for designers, even if their performance would not be considered “high resolution” within haptics research. We treat them as basic examples. Beyond these, the haptics community has explored a wide range of alternatives, such as shape-memory alloys \cite{messerschmidt_ANISMA_22}, voice coils \cite{culbertson2018social}, hydraulics, electrostatic actuators \cite{shen2023fluidreality, leroy2020multimode, youn2025skin}. \revise{Different actuators with custom mechanisms can also render haptic feedback across diverse modalities, such as altering temperature \cite{Mazursky2024thermal}, inertial perception \cite{nith2023jumpmod}, softness \cite{tao2021softness}, and stickiness \cite{mazursky2024stick}. Together, these approaches point to a rich design space for exploring expressive robotic touch.}

While pneumatics are often framed in soft robotics as bio-inspired “artificial muscles,” they need not replicate muscle movement when applied to touch. In Blo-nut, air actuation plays a central role in shaping perceptions of aliveness. Its profound, smooth choreographies exemplify the \textit{\textbf{animation}} strategy, as their rhythmic expansions and contractions are readily associated with breathing and lively movement. At the same time, its hybrid form and mechanism—half rigid and half soft, with three independently controlled pneumatic chambers—was intentionally designed with asymmetry to elicit a sense of otherness.
The Pelvic Chair also employs pneumatics, but here the air pockets expand against the constraints of the seated body. The design intentionally used elastic shape-change actuators, as they could follow the body’s contours while creating sensations of being softly held or embraced, rather than poky, which could be pointed, hand-like, or intrusive on the intimate area.
The placement, scale, and shape of these pockets are carefully attuned to the pelvic floor’s anatomy and feelings through \textbf{\textit{somatic grounding}}. This is a new strategy we term to describe how soma design \cite{hook2018designing} practices can be used to explore the felt qualities that make a robotic touch experience “feel good” or “feel right.” Unlike strategies that work by unfamiliarizing familiar touch drawn from another subject, somatic grounding relies on design judgment rooted in the designer’s first-person felt experience, ensuring that robotic touch resonates with somaesthetic awareness and meaning. Through the precise mapping of inflatable pockets to pelvic-floor muscles, the touch \textbf{\textit{exaggerates}} the sense of these muscles, letting the sitter feel them “closer” by feeling where they are, how they tense and release, and how they connect with one another and with the rest of the body. Here, otherness helps to estrange the sitter’s body, fostering deeper touch awareness and engagement. 

TactorBots use micro servo motors to drive tactors. While many modules were inspired by hand gestures, the strategy of \textit{\textbf{minimalism}} guided their 1 DoF mechanisms. At the same time, the servos’ inevitable jiggling, whirring, and visible parts continually reminded users of their robotic character. The Rub and Stroke tactors, which employed a spinning contoured barrel with raised dots to create lateral feedback, attracted particular interest. By \textit{\textbf{abstracting}} the touch behavior to focus only on perceptual cues, these tactors enhanced otherness, showing how a robotic interface can evoke relatable touch without reproducing the way humans deliver it with their hands. 
Shape-Kit is unusual in that it employs no robotic actuator. Instead, it is analog, actuated directly through hand manipulations transmitted by long, flexible cables. Designers can craft it with bare hands or hand-held props, and the mechanism translates these familiar actions into re-scaled, pin-based sensations. This process \textbf{\textit{exaggerates}} the fine sensations of hand touch into sharpened, lower-resolution pin movements. The \textbf{\textit{juxtaposition}} between hand actuation and pixelated output brings its otherness to the surface. 

\textbf{\textit{Movement \& Control.}} Movement defines the dynamic feasibility of robotic touch, while control determines how those movements are authored and performed spatially and temporally. We all know that it is challenging for robotic interfaces to move with the same dexterity as our hands. But from another point of view, this also means robotic interfaces can move in ways impossible for living organisms, where their otherness emerges and provokes us to interpret them with broader imagination.

In TactorBots, each module is driven by a single servo, constraining movement to axis-aligned rotation in 1 DoF. These movements are \textbf{\textit{abstracted}} into control parameters, such as minimum and maximum range, movement duration, and pause time at each end, authored through a web interface with sliders. The resulting movement is rendered at a constant speed in continuous, precisely repeatable loops. When optional randomness is introduced, participants noted that it added a sense of robotic liveness, though the touch was still perceived as distinctly mechanical. 
Shape-Kit is similarly limited to 1 DoF, with spring-back pins that only move up and down. This filtering \textbf{\textit{distorts}} dexterous hand behavior to a single dimension, while also transforming them in scale and texture, resulting in a new type of dynamically natural yet other touch. Within this constrained space, designers could concentrate on pressure patterns, rhythm, body placement, and texture without being overwhelmed or distracted by unrestricted options. 
Blo-nut’s pneumatic chambers are controlled through a music-jamming interface, allowing its shape-changing to be choreographed like a dance. Rather than referencing limbs or bodily gestures, the three chambers inflate and deflate in sequence, aligning with the strategy of \textbf{\textit{animation}} to foreground visual expressiveness, which could also be tactilized with relatable meaning. 

The Pelvic Chair combines pneumatic actuation with the constraints of the seated body. Employing the \textbf{\textit{somatic grounding}} strategy, its movement sequences were shaped through iterative soma design sessions, from early low-fidelity prototypes with manual pumping to later high-fidelity interfaces controlled by a MIDI editor. The touch of its pneumatic pockets was distinctive: the sitter’s weight and body shape guided the air to fill gaps or areas with less pressure, producing a “filling in” sensation unlike any living organism’s touch. Drawing on the experiential quality of relaxational power, for example, soma designers emphasized a sensation not of “being touched” but of “being released.” When air pockets deflated after a held inflation, muscles sank back into the chair, creating a deep sense of relaxation. The final actuation sequence was carefully orchestrated to progress from the less intimate lower back and inner thighs toward the pelvic floor. Movements were deliberately slow, gentle, and prolonged, with some repetition to respect the slowness of visceral interpretation. The entire touch experience was found to support improving bodily awareness and meaning-making.

Across these systems, otherness is embedded in movement and control through robotic systems’ inherent constraints and programmability. Constrained DoF reduces the complexity of mimicking human touch, while programmable control enables precision, repetition, and extreme slowness or fast speed that living bodies, limited by energy and fatigue, cannot easily achieve.

\subsubsection{Intertwining Strategies and Attributes}

Some strategies can be effective because they apply across multiple attributes at once, impacting the touch experience holistically. Consider the case of \textbf{\textit{transparency}}. While only TactorBots employ a fully transparent top enclosure, both Shape-Kit’s long cable tubes and Blo-nut’s silicone skin are semi-transparent, allowing the mechanism to be partially revealed. On one hand, transparency is an optical property of material choice; on the other hand, it powerfully influences the visually perceived form. Its primary value lies in making mechanisms and actuator movements legible, which in turn supports trust building and enhances the sense-making of touch. 

Another typical example is \textbf{\textit{juxtaposition}}, which highlights the value of encountering otherness through contrasts between the familiar and the unfamiliar \cite{boer2018reconfiguring_blo-nut, zhou2023tactorbots}. As shown in Figure \ref{fig:deconstruction}, juxtaposition often sits at the intersection of multiple attributes, or is achieved in combination with other strategies. For instance, Shape-Kit relies on minimalism in form, together with the distortion of direct hand manipulation. The Pelvic Chair, by contrast, draws on the tension between the objectness of an ordinary stool and the orchestrated touch that exaggerates awareness of the intimate seated body. Juxtaposition here becomes a powerful way of shaping otherness with enhanced ambiguity, which in turn invites curiosity and deeper engagement.

In revisiting our four cases, we analyzed where otherness can be embedded and how it can emerge or be shaped during the design of robotic touch systems. We offer these insights as takeaways, while recognizing that many more strategies and patterns remain to be uncovered as the community continues to explore robotic interfaces in new forms, with diverse actuators and in varied contexts.

\begin{figure*}[ht]
  \includegraphics[width=\textwidth]{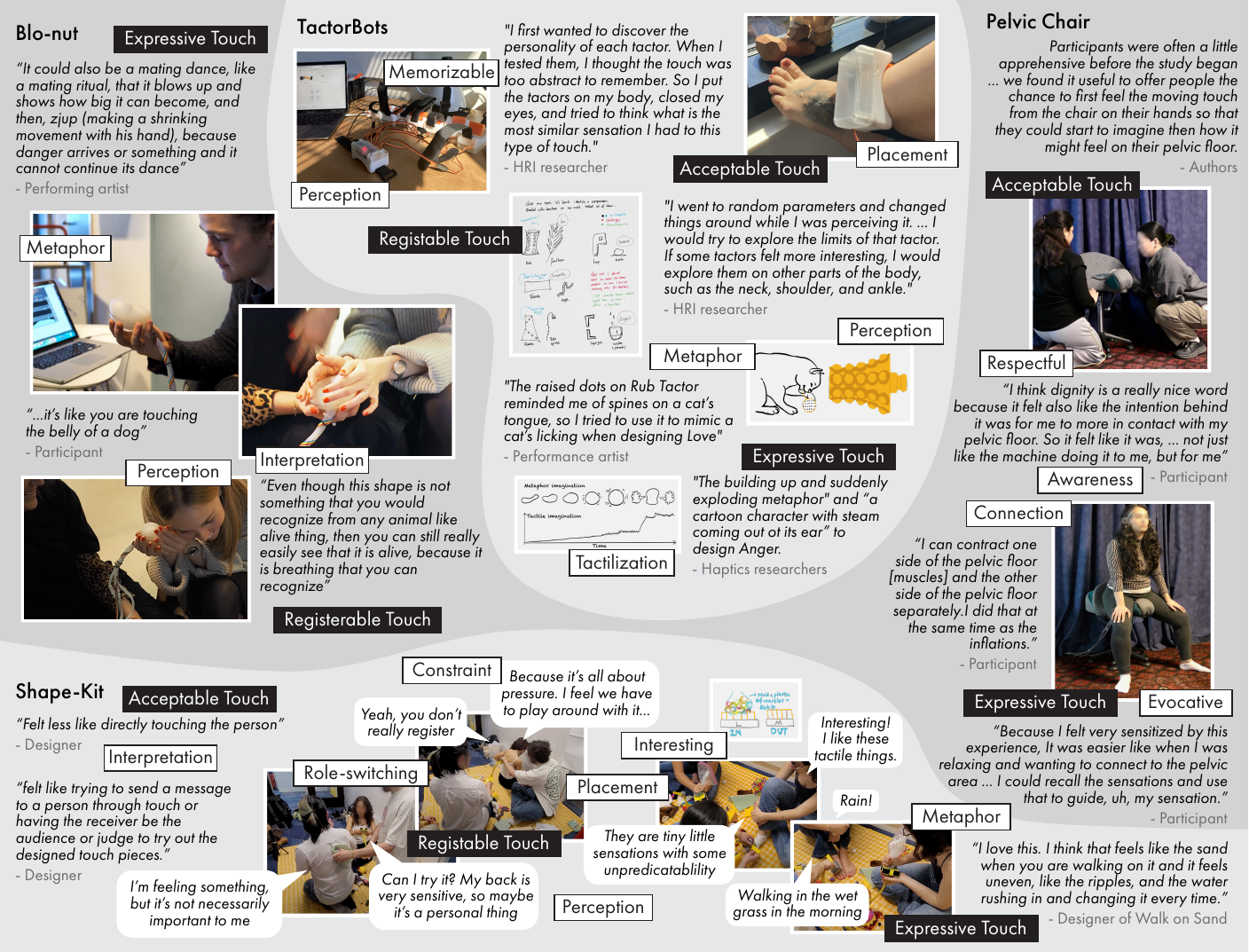}
    \caption{Selected quotes from each design research case illustrating participants’ experiences of engaging with expressive robotic touch. We annotate key experiential themes (white background) and highlight dimensions of robotic touch meaning-making (black background).}
  \Description{Figure presenting selected participant quotes from four robotic touch case studies—Blo-nut, TactorBots, Shape-Kit, and Pelvic Chair—annotated to illustrate experiential themes. Quotes describe interpretations such as breathing, mating dances, animal metaphors, cartoon imagery, bodily awareness, dignity, and playfulness. Annotations highlight dimensions of touch meaning-making: acceptable, registerable, and expressive touch.}
  \label{fig:sense-making}
\end{figure*}

\subsection{Why Otherness Matters in Touch Meaning-Making}
\label{otherness_in_sense-making}

Our case studies span diverse modes of exploration and engagement. Blo-nut \cite{boer2018reconfiguring_blo-nut}: experts engaging in choreographic sketching to explore robot expressions and designers interpreting the robot’s behavior to imagine new relationships; TactorBots \cite{zhou2023tactorbots}: designers creating expressive touch and testing it on their own bodies; Shape-Kit \cite{zhou2025shape-kit}: groups of designers collaboratively prototyping and iterating through role-switching; and Pelvic Chair \cite{yadav2025somatic_literacy}: soma design researchers carefully orchestrating touch sequences and then be experienced by participants. Each case offers rich qualitative analysis of these design and sensorial explorations, providing valuable insight into how people make sense and make meaning of robotic touch. \revise{Revisiting these accounts through reflective RtD analysis, we \camera{as the original design researchers of the systems who conducted those empirical studies} identify three recurring dimensions of touch meaning-making: acceptability, registerability, and expressiveness. These dimensions emerged from cross-case reflection grounded in long-term engagement with the systems—as designers, researchers, and, at times, participants—based on design iterations, accumulated observations, and experiential judgments over time \cite{redstrom2017making}.} Drawing on quotes highlighted in each work, we curated a portfolio (Fig. \ref{fig:sense-making}) that annotates experiential themes in relation to these dimensions. Here, we examine \textbf{\textit{why}} otherness matters in expressive robotic touch design by investigating the role it plays across them.

\subsubsection{Otherness for Acceptable Robotic Touch}
\revise{A prerequisite for making sense of robotic touch is to accept the robotic surface touching your skin}. In interpersonal contexts, touch is deeply tied to cultural context and social norms \cite{suvilehto2015topography}. When extended into touch technologies, prior work on mediated social touch, where robotic interfaces deliver touch on behalf of a distant person, has shown that sender–receiver’s genders and relationships are critical factors \cite{smith_communicating_2007, ipakchianaskari2021receiving, askari2022separated}. Touch from humanoid robots has previously been judged more acceptable than from a human stranger, but largely because robots were perceived as lacking intention or emotion \cite{hoffmann2017robot}. In contrast, the cases we discuss show how robotic interfaces with qualities of otherness, even when highly expressive, sometimes seemed to gain an exemption from these norms. Participants showed high acceptance in engaging with the touch, openness in exploring bodily boundaries (Fig. \ref{fig:sense-making}), and, notably, no explicit concerns about gender or the touch initiator’s social relationship.

One reason lies in contextual otherness: none of the systems were prescribed as representing another person, so touch was interpreted as originating from the device itself, an “other subject” in its own right. Even in Shape-Kit, which directly transduced hand manipulations through analog cables, the resulting sensations were perceived as design outcomes rather than remote contact from the crafter (Fig. \ref{fig:sense-making}).
The \camera{system's otherness} also contributed to acceptability. These interfaces were intentionally designed in non-human, non-hand-like forms and textures. When touched by a human hand, even with eyes closed, one can easily imagine the whole body that the hand belongs to \cite{zhou_emotitactor_22}. In contrast, when touched by a robotic interface whose sensations feel other, the imagined “toucher” is more ambiguous and detached from a specific social identity. 

Trust is another crucial factor of acceptability. Blo-nut, TactorBots, and Shape-Kit used transparency and minimalist mechanisms to make their behavior understandable and foster trust-building. In the Pelvic Chair, trust was especially critical given the intimate body area involved. While its system- and context-level otherness contributed, its acceptability relied heavily on the careful study facilitation. This included providing a private and relaxed environment; priming participants by letting them closely observe the chair’s behavior and first try the touch on their hands to learn what they should expect. Importantly, acceptance of the system does not imply acceptance of all its touches. In the Pelvic Chair, ultimate acceptability depended on delicately orchestrated touch movements and sequences, guided by the designers’ first-person perception. Together, these considerations made the interaction feel not only acceptable and safe but also dignified and enjoyable.

\subsubsection{Otherness for Registerable Robotic Touch}

Human skin is constantly stimulated, yet most touches in daily life are filtered out to reduce cognitive load. For robotic touch to matter, it must be registerable: not only perceptible, but also interesting and memorable (Fig. \ref{fig:sense-making}). Our analysis suggests that what counts as registerable is highly individual-dependent, often judged through bodily sensibility and intuition rather than logical or theoretical reasoning. Here, otherness plays a role by eliciting \textbf{\textit{alternative perceptions}}—producing touches that carry a sense of unfamiliarity, disrupting habituation, and prompting attention and curiosity.

Noticeability depends partly on thresholds of skin sensitivity, which vary across body locations and individuals \cite{lederman2009tutorial}, but also on whether the body is prepared to attune attention to tactile sensation. While such conditions are subjective, some methods, such as soma design \cite{hook2018designing}, offer practices for sensitizing the body to foster the attentional readiness, which has been used in the Shape-Kit and Pelvic Chair cases.

However, making touch perceptible does not guarantee that it is registerable. In Blo-nut, TactorBots, and Shape-Kit, designers often relied on sensory exploration to discover what felt interesting. It was typically easier for them to articulate when a touch felt \textit{“not right”} than to explain what made it \textit{“right,”} with judgments guided by bodily intuition. Such intrigue was often sparked by alternative perceptions elicited through otherness, which provokes curiosity, while people also often needed a conceptual handle to anchor the experience to make it memorable. For instance, in TactorBots, the Rub Tactor, whose touch was the most surprising, gained the most interest, while participants shared their impression by associating it with metaphors such as spines on a cat's tongue. Another designer in TactorBots experimented with parameters across body sites to explore the \textit{“personality”} of each tactor, eliciting metaphors such as \textit{“spikes of a rose”} or \textit{“a butterfly trapped in a cocoon.”} Such generative metaphors helped the designer recall and work with sensations as design materials. 

The Pelvic Chair shows how even entirely novel touches can become registerable when carefully orchestrated. Its tactile sensations were initially “too other” to connect with any associations from personal history, both because of the system design and the unusual location of touch. Yet by gradually sequencing from the less sensitive back and inner thigh toward the pelvic floor—beginning as barely registrable and progressing to stronger inflations—the system guided participants’ attention until the touch became noticeable and focused. Crucially, because the shapes and placements were attuned to pelvic anatomy, the inflatables stimulated distinct muscle regions. This means participants could recall these experiences by actively moving or finely controlling their pelvic muscles during and even after the session, making it memorable.

\begin{figure*}[ht]
  \includegraphics[width=\textwidth]{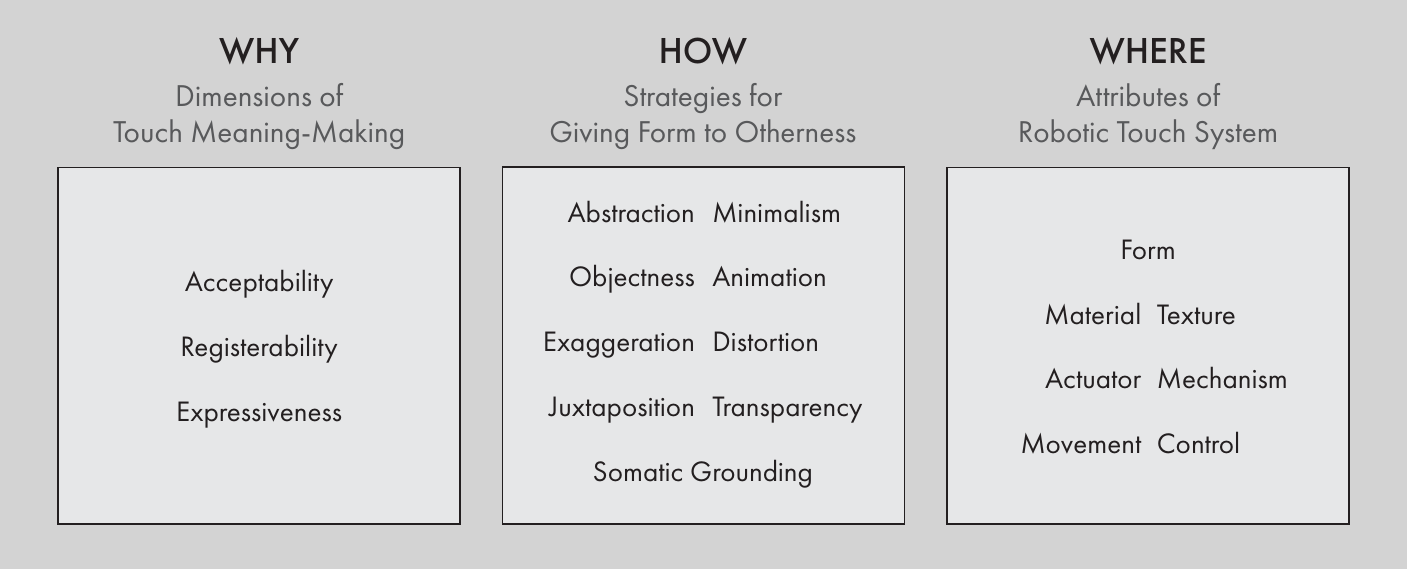}
 
    \caption{Overview of our developed design languages as handles to help designers think and work with otherness in practices. It involves: Why-dimensions of touch meaning making, How-strategies in giving form to otherness, and Where-attributes in robotic touch system.}
  \Description{Figure includes three columns. In the left column: Why; dimensions of touch meaning making; acceptability, registerability, and expressiveness. In the middle column: How; strategies in giving form to otherness; Abstraction, Minimalism, Objectness, Animation, Exaggeration, Distortion, Juxtaposition, Transparency, Somatic Grounding. In the right column: Where; attributes in robotic touch system; Form, Material, Texture, Actuator, Mechanism, Movement, Control}
  \label{fig:overview}
\end{figure*}

\subsubsection{Otherness for Expressive Robotic Touch}

The expressiveness of robotic touch centers on the receiver’s interpretation. Prior research has shown that the meaning-making of artificial touch is highly contextual and subjective \cite{foo_user_2021, kim_swarmhaptics_2019, price_2022_making_meaning}. Even in interpersonal contexts, psychology studies often rely on forced-choice methods that map touches to predefined meanings such as emotions \cite{hertenstein_touch_2006, hertenstein_communication_2009, thompson_effect_2011}. While such studies demonstrate that touches can be distinguishable or decodable, they also highlight the difficulty of establishing a generalizable language of touch without specialized training. Rather than treating this as a limitation, we argue that robotic touch does not need to convey accurate information, emotion, or affect for it to be expressive. Following design research that embraces openness and multiple meanings \cite{sengers2006interpretation}, we suggest that expressive robotic touch should not aim for a designer-assigned, single “correct” interpretation, but rather be understood as a generative resource that allows multiple and alternative interpretations to emerge and co-exist. Here, the ambiguity elicited by otherness plays a generative role, provoking these \textbf{\textit{alternative interpretations}}.

From the touch designer’s perspective, otherness supports creative imagination. In Blo-nut, TactorBots, and Shape-Kit, designers were invited to ideate and author expressive touch experiences. Because the systems and their sensations carried qualities of otherness, designers could not simply draw on familiar touch memories. Instead, they consciously made sense of the sensations through metaphors and associations, which also became the way they described their design ideas (Fig. \ref{fig:sense-making}). Designers approached this in different ways: some began with a concept they wanted to realize, while others engaged in open-ended sensory exploration that allowed ideas to emerge. Across cases, we found that design ideas, metaphors, and touch perceptions were closely intertwined; each could spark the others, prompting designers to shift ideas, discover new metaphors, or tune sensations. While Blo-nut was framed as a social robot and thus sometimes associated to entities based on the system, such as\textit{ “a church organ”} or \textit{“a sea creature,”} most metaphors in the cases described the qualities of the touch itself (e.g., \textit{“cat’s tongue,” “rose spikes,” “walking on sand”}) rather than imagining a character or social “toucher” that performed all those sensations, which might constraint the imagination. 

From the audience’s perspective, otherness evokes personal interpretation. Even though Blo-Nut was staged as a prototype for a future pet, it was not straightforward for the groups of designers to conceptualize the robots. They could relate it to domestic animals, but only partly, like \textit{“touching the belly of a dog,”} as its form and behavior could not match the familiar pet in their expectation. Instead, the robot was considered as an \textit{“alive machine”} which was seen as a gadget, product, or laboratory prototype, that could be switched on and off, while at the same time, its breathing-like actuation let them to \textit{“easily see that it is alive,”} provoking emotional responses such as \textit{“you are upsetting it,” “don’t kill it,”} or \textit{“it likes me, I was calming it.”} 

The Pelvic Chair shows how meaningful touch does not need to speak to the mind but can communicate directly with the body. Like music or fragrance, its expressiveness could be grounded in sensation itself rather than in metaphor or explicit language. When experiencing the Pelvic Chair, participants found the touch too new to connect with personal history or metaphors; however, they appreciated its aesthetic qualities, the feeling of deep relaxation, and finer motor control over those large and small, often unnoticed pelvic muscles. They described the chair as \textit{“not doing something to them, but for them,”} using subtle touches to help them notice, connect, and engage more deeply with their own bodies. Some said it allowed them to \textit{“breathe through the pelvic floor”} or that \textit{“the pelvic region feels closer in the body.”} Rather than passively receiving, some actively engaged by delicately controlling their muscles to play and move in correspondence with the chair’s touch, an experience they felt would not be possible with human touch. These meaningful sensations contributed to somatic literacy, leaving participants with embodied touch memories that deepened their awareness of their own body’s expressiveness.

While touch interpretation is always subjective, we observed that expressive robotic touches created by one or a group of designers could still resonate with others, even if not universally \cite{stahl_validity_21}. The Pelvic Chair illustrates that, besides registerability, felt aesthetics identified through soma design could also be appreciated by participants beyond the original designers. Shape-Kit, which enables collaborative touch design, further demonstrated shared meaning-making. As shown in Fig. \ref{fig:sense-making}, one designer noticed an intriguing sensation without knowing its meaning, while after switching roles, the other immediately associated it with \textit{“rain,”} which the group extended into a shared metaphor of \textit{“walking in wet grass in the morning.”} At times, an idea was rooted in one designer’s personal history, who then guided the team to explore sensations that matched that memory while sharing the story behind it. Through this process, novel shared touch memories might emerge. We see the interpretation of expressive touch from robotic interfaces with otherness as comparable to how people interpret artworks: an artist may provide a title or description to share their story and seek resonance, while at the same time, the audiences are invited to appreciate it while having their personal readings.

\revise{
\subsection{Overview}

 



In Fig. \ref{fig:overview}, we present an overview of the design languages articulated through our reflective analysis across cases. Specifically, we identify three dimensions of touch meaning-making, nine strategies for giving form to otherness, and seven attributes for otherness to be embedded in the robotic touch interfaces. The overview is organized around the questions why, how, and where, mapping a flow from design intention or motivation, through design action, to points of implementation. We deliberately do not add any arrows between them, treating these design languages as conceptual handles that invite open-ended design exploration with flexibility and fluidity.

We offer this overview as a reflective resource \cite{lenz2013exploring} to support designers in thinking with, talking about, working through, and reflecting on designing expressive robotic touch with otherness. We want to emphasize that this overview is not an exhaustive list. Rather, it shows shared experiences and insights developed through sustained RtD practices and reflection by \camera{us, the designers of the cases}. We expect that additional vocabularies will continue to emerge as the community further explores otherness as a design quality in expressive robotic touch and related domains.

}

\section{Risks and Tensions}

While we have primarily argued for the value of otherness as a quality in designing expressive robotic touch, working with it also raises tensions and risks. 

\textbf{\textit{Otherness does not equate to low fidelity.}} While these design research cases may appear low resolution compared to emerging haptic technologies, they were executed with care and precision. Strategies like minimalism and abstraction are not shortcuts to easy or cheap design but deliberate pathways that evoke distinct aesthetic qualities and gestalt delight. Just as pixel art and low-poly aesthetics continue to flourish alongside ultra-high-resolution displays and VR environments, otherness likewise deserves exploration within high-resolution haptics. It is not opposed to technological progress but a constant reminder of the beauty of alternative values: using ambiguity, distortion, and reduction to expand the expressive possibilities of touch.

\textbf{\textit{Tension remains in the exploratory design and final implementation.}} Cases like Blo-nut, TactorBots, and Shape-Kit are experimental artifacts or toolkits where otherness can be productively harnessed to probe what expressive robotic touch could be. Yet when translated into future everyday end-user devices, its qualities may shift: does otherness only reside in the system, or can it also be stored and carried in the touch itself? For example, would a touch crafted in Shape-Kit evoke similar interpretations if rendered on a compliant on-skin pin array? Whether otherness is a transferable quality remains an open question.

\textbf{\textit{Acceptability of touch with otherness also raises the risk of over-trust.}} In our discussed cases, otherness often appeared to exempt robotic touch from established social norms, opening broader design opportunities. Yet most of these touches were created and initiated by designers themselves, or explored under carefully controlled conditions. In real-world deployments, however, questions of who initiates the touch, when, where, and under what circumstances may again become crucial. Otherness should not be an excuse to bypass such considerations. As robotic touch systems increasingly collaborate with AI, new ethical questions around consent, agency, and the emergence of novel social norms will demand serious attention \cite{sassmannshausen2023human}.

\section{Conclusion}

This paper draws on HRI theory and design research to argue for embracing robotic touch’s inherent otherness as a design quality rather than a limitation, intervening in the replication-driven mindset that dominates current haptics and robotic touch research. Rather than criticizing current trends, we focus on exploring alternatives—asking what else robotic touch can do, especially in expressive and everyday scenarios. We ground this argument theoretically while also analyzing precedents in art and design and design research cases in HCI to develop shareable insights and design languages. By unpacking four cases in depth, we examined where and how otherness is embedded, and why it matters for touch sense-making. At the same time, we propose tensions and risks on fidelity, transferability, and trust that need to be carefully negotiated if otherness is to become a sustainable quality in future robotic touch design.

\section{Acknowledgments}
The authors extend their gratitude to our collaborators in the design research cases. We thank the reviewers for their constructive feedback. We also thank all friends, colleagues, and professors who have provided feedback and support throughout this long-term articulation journey, especially Yujie Tao, Yiran Zhao, Jasmine Lu, Andreas Lindegren, Laia Turmo Vidal, Alice Haynes, Kristina Höök, Harpreet Sareen, Anna Vallgårda, Laura Devendorf, Wendy Ju, and Patricia Alves-Oliveira.

Co-funded by the European Union (ERC, Intimate Touch, 101043637). Views and opinions expressed are however those of the author(s) only and do not necessarily reflect those of the European Union or the European Research Council. Neither the European Union nor the granting authority can be held responsible for them.



\bibliographystyle{ACM-Reference-Format}
\bibliography{reference}

\end{document}